\documentclass[prl,onecolumn,superscriptaddress,nofootinbib]{revtex4-1}
\usepackage{graphicx}
\usepackage{color}
\usepackage{amsmath}
\usepackage{amsfonts}
\usepackage{mathrsfs}

\begin{document}

\title{Quasiperiodic Frank-Kasper phases derived from the square-triangle dodecagonal tiling }

\author{Jean-Fran\c cois Sadoc}
\email{jean-francois.sadoc@u-psud.fr}
\affiliation{Laboratoire de Physique des Solides (CNRS-UMR 8502), B{\^a}t. 510, Universit{\'e} Paris-sud Paris-Saclay, F 91405 Orsay cedex, France}

\author{ R\'emy Mosseri }
\email{remy.mosseri@upmc.fr}
\affiliation{  Laboratoire de Physique Th{\'e}orique de la Mati{\`e}re Condens{\'e}e, UPMC, CNRS UMR 7600, Sorbonne Universit{\'e}s, 4 place Jussieu, F-75005 Paris, France}

\begin{abstract}
Frank-Kasper (F-K) phases form an important set of large-cell crystalline structures describing  many inter-metallic alloys.  They are usually  described in term of their atomic environments, with atoms having $12, 14, 15$ and $16$ neighbours, coded into  the canonical $Z_p$ cells (with $p$ the coordination number), the case $p=12$ corresponding to a local icosahedral environment. In addition, the long range structure is captured by the geometry of a network (called either ``major skeleton'' or ``disclination network'') connecting only the non-icosahedral sites (with $p\ne 12$). Another interesting description, valid for the so-called ``layered F-K phases'', amounts to give simple rules to decorate specific periodic 2d tilings made of triangles and squares and eventually get the 3d periodic F-K phases. Quasicrystalline phases can sometime be  found in the vicinity, in the phase diagram, of the F-K crystalline alloys; it is therefore of interest to understand if and how the standard F-K construction rules can be generalized on top of an underlying quasiperiodic structure. It is in particular  natural to investigate how well square-triangle quasiperiodic tilings with dodecagonal symmetry, made of square and (equilateral) triangles, can be used as building frames to generate  some F-K-like quasicrystalline structures. We show here how to produce two types of such structures, which are quasiperiodic in a plane and periodic in the third direction, and containing (or not) $Z_{16}$ sites. \textit{It is a pleasure for us to contribute to a  special issue of ``Structural Chemistry'' in honour of Alan Mackay, for his 90th birthay. Alan's work has been a source of inspiration for most of the researchers working in the field of complex atomic structures.}

To appear in ``Structural Chemistry''.

\end{abstract}

\maketitle

\section{Introduction}
\label{intro}

\begin{figure}[tc]
\includegraphics  {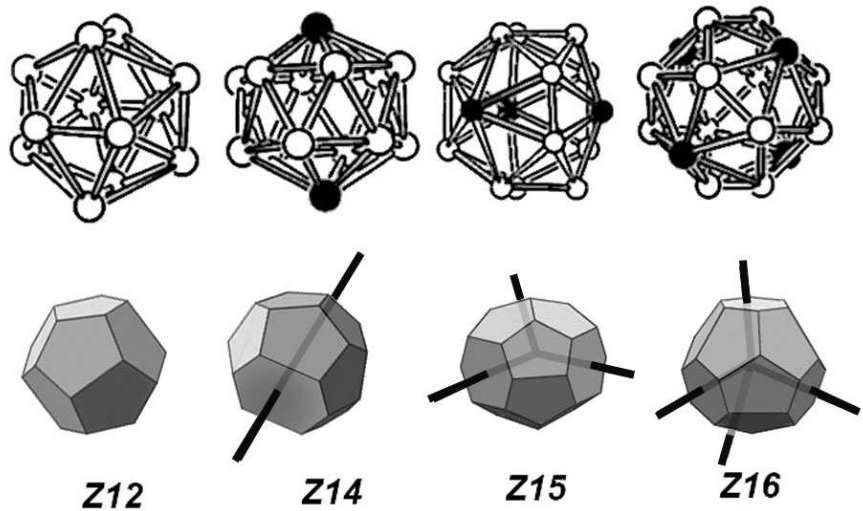}
\caption{\label{fig:f1}The four main Frank-Kasper coordination polyhedra (top figure)  $Z_{12}$,  $Z_{14}$,
 $Z_{15}$ and $Z_{16}$, represented also by their Voronoi cells (lower figures). Frank-Kasper lines (or disclination lines), shown as black lines, start from the center of the coordination polyhedra and run through the black sites, or from the Voronoi cell centres and go through hexagonal faces, defining a ``major skeleton'' in the structure.}

\end{figure}

Beside simple metallic systems found with b.c.c and f.c.c structures, with few atoms per unit cells, many intermetallic alloys display more or less intricate structures with polytetrahedral type of packing, and possibly a large number of atoms in the unit cell, like in the so-called Frank-Kasper (F-K) phases\cite{frankkasper}. It is standard in that case to focus on the local atomic environments, the canonical $Z_p$ cells (see fig.~\ref{fig:f1}), the case $p=12$ corresponding to a local icosahedral environment.

 A long-standing question in quasicrystal studies is to relate the atomic order present in quasicrystalline phases with that of F-K phases, some of which being identified as rational approximants of the former; dually, these quasicrystalline phases (close in the phase diagram) could then be thought as mere quasperiodic generalizations of the periodic F-K phases.

%
%

As will be detailed below,  F-K structures can also be generated with an atomic decoration procedure applied on top of a plane tiling made of triangles and/or squares, building simple atomic layers. To go further, one should then understand how the F-K construction rules can be accommodated with an underlying quasiperiodic tiling; it was therefore natural to investigate how well dodecagonal quasiperiodic tilings, made of squares and (equilateral) triangles, can be used as building frames to generate  some F-K-like quasicrystalline structures.  This is the main task of the work presented here.

 Note that such tilings have been used to model dodecagonal quasicrystals found in metallic alloys related to Frank-Kasper  phases \cite{ishimasa,chen} and, more recently, in dendrimeric supramolecular liquid crystals \cite{zeng}.
 We show here how to produce two types of such structures, which are quasiperiodic in a plane and periodic in the third direction, and containing (or not) $Z_{16}$ sites (sites with 16 neighbours).

In a first step, we detail the F-K rules for periodic systems. Four parallel atomic layers are disposed with respect to a periodic planar frame made of triangle and squares. The occurrence of $Z_p$ sites and periodic disclinated networks are direct  consequences of these rules.

In order to generate quasicrystalline F-K-like structures, we then start with a particular tiling with square and triangles, the dodecagonal square-triangle quasiperiodic tiling. We first recall a standard method to produce the 12-fold symmetric quasiperiodic tiling,  the inflation/deflation approach (see ref \cite{stampfli,hermisson}), for which we propose a new simplified algorithm. Notice that a standard cut and project approach from a higher dimensional space is also possible, but operationally complex because the acceptance domain (used to select the tiling points) has a fractal shape in that case.

We then return to the question of  atomic decorations leading to close-packed configurations, and extend the F-K rules to this wider quasiperiodic domain. A particular such decoration, corresponding to the type of atomic order found in the crystalline F-K $\sigma$ phase, was already used by Gahler \cite{gahler88} to model the first dodecagonal quasicrystalline alloys found in Ni-Cr, V-Ni and V-Si-Ni alloys. Notice a slight difference in that case since the underlying quasiperiodic tiling contained thin rhombuses in addition to the square and triangles (in that case the acceptance domain is a regular dodecagon instead of a fractal shape, making the construction simpler). See also ref. \cite{mihalwidom} for further developments. As for the periodic case, four  atomic layers are generated, with atomic positional order following that of the underlying quasiperiodic tiling, these quasicrystalline atomic layers being then repeated periodically. In the present work, we discuss  two different decoration schemes to be applied on the square-triangle quasiperiodic tiling, leading to two types of $Z_{15}$ sites (in-plane or connecting different planes) as well as a variable amount of $Z_{16}$ sites.

Beside the perfect 12-fold quasiperiodic tiling, a rich set of square-triangle tilings can also be derived, with either lower (hexagonal) symmetry or even forming random tilings. From any such tiling, an atomic decoration procedure could be applied, locally inspired by the F-K rules, leading to a large set of tetrahedrally close packed (TCP) structures. Among these generalized structures, we shall discuss an interesting new set which can interpolate between the F-K $Z$ phase, whose underlying tiling is a triangular one, and the $A15$ phase with a square underlying tiling. This set is periodic along two directions, with, along the third one, a sequence of triangle and square rows which can take any form (periodic, quasiperiodic, disordered). An explicit quasiperiodic example is described in the appendix.

\section{Frank-Kasper phases}

F-K structures are tetrahedrally close-packed periodic structures, first described by Frank and Kasper \cite{frankkasper}, and later studied extensively by many authors, among which notably D. and C. Shoemaker \cite{shoemaker}. The  atoms occupy (so-called) $Z_p$ sites, where $p$, the site coordination number, is equal to either $12, 14, 15$ or $16$ (fig.~\ref{fig:f1}), and have interpenetrating triangulated coordination polyhedra;  the $Z_p$ sites Voronoi cells  have  $12$ pentagons (the $Z_{12}$ Voronoi polyhedron is a pentagonal dodecahedron) and $p-12$ hexagons.

When icosahedral quasicrystals were discovered in intermetallic compounds, the large cell structures found nearby in the phase diagram became of parallel interest, as they were now viewed as periodic approximant of the perfect quasicrystalline structure. The metallurgist community quickly  considered that Frank-Kasper-like phases \cite{frankkasper,shoemaker} or related phases (like $\alpha$-Mn or CaCu$_5$ for instance) would be   good models for these large cell periodic structures. Indeed,  they already present a high degree of local and medium range icosahedral configurations, like for instance in the Bergman structure \cite{bergman57}. Quite rapidly, decagonal and dodecagonal quasicrystalline structures were also discovered and further related to F-K structures. More recently, dodecagonal quasicrystals have also been found  in soft matter systems, like in dendrimeric supramolecular liquid crystals \cite{zeng,ungar}, a particularly interesting discovery showing that metallic bonding is not a prerequisite for stable quasiperiodic order. Note in addition that F-K phases were also  observed in micellar structures \cite{hajiw}.

%
\begin{figure}[tbp]
\includegraphics [width=12cm]{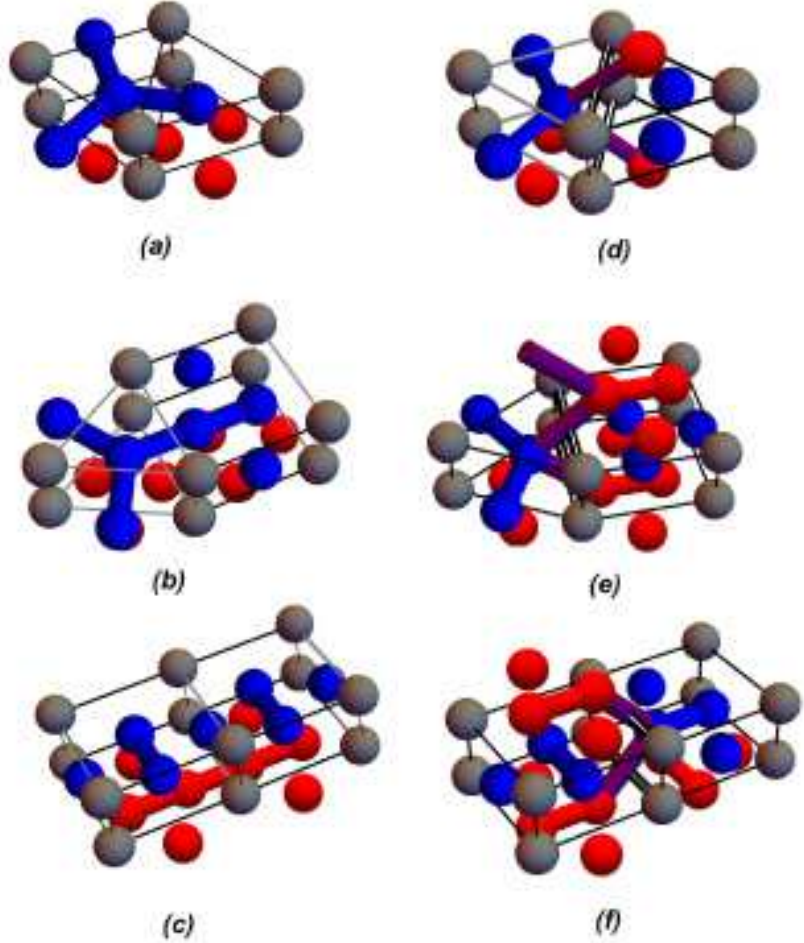}

\caption{The two decoration modes leading to $Z_{16}$-free structures  (left part), or with $Z_{16}$ sites (right part), for the three types of tile interface : triangle-triangle (a,d), triangle-square (b,e) and square-square (c,f).
Let us first describe the left part (a,b,c). Sites located on the secondary layers at height $1/4$ and $3/4$ sit at the tile vertices and are shown in grey. We aim to discuss the decoration of  primary layer (at height $1/2$)  with atoms located inside the triangular and/or square prisms formed by secondary layer sites. As explained in the text, there are two sets of edges (marked as dark or light-grey edges), with respect to their angular value (modulo $\pi/3$). Edges of a given triangle will belong to one of the two sets, while adjacent edges of a square will alternate among the two sets. There are two different primary layer decoration rules (called rules $I$ and $II$), at the vertical  of  a triangle of the secondary layer, depending on the triangle colour, and whether the primary layer is at height $0$ or $1/2$. This point will appear more clearly in fig.~\ref{fig:decoz15}.
 The figure displays rule $I$, on the primary layer at height $1/2$, with atoms inside triangular prisms (with the triangles being of light-grey colour), or square prisms. These atoms are shown in blue (colour on line), with the piece of disclination network to which they belong. They fall in the middle of the triangles, in the middle of the dark grey edges and  displaced from the middle of the light grey edges by 1/4 of the square edge length. Rule $II$ applies when the triangles have dark edges, in which case the decoration adds atomic sites on top of the edge centres. Sites located on primary layers at height $0$ are also shown (in red with colour on-line). Note that these two rules for triangle decoration drive the decoration on top of  the squares, with atomic sites on top of dark edges and slightly displaced for the light ones.
  Sites at height $1/2$ in the triangular prism middle are $Z_{15}$ sites, connected to $Z_{14}$ sites through a triangle-square interface. Sites falling on the dark grey mid-edge are $Z_{12}$ sites. Finally, grey sites on the secondary layers are all $Z_{14}$ sites, connected by vertical disclination lines.
 Now, for the primary layer at height $0$, the decoration is similar, but with rules $I$ and $II$ reversed with respect to the edge colouring.
Let us turn now to the layer decoration with $Z_{16}$ sites, shown on the right part (d,e,f). As said in the text, some edges are now ``double edges'' separating stripes. Across these double edges, rules $I$ and $II$ are reversed. This leads to connecting the disclination lines between two primary layers (in blue and red on-line). A formerly $Z_{15}$ (resp. a   $Z_{14}$) site is turned into a $Z_{16}$ (resp. a   $Z_{15}$) site.}
\label{fig:discli_z15_z16}
\end{figure}

\subsection{Major skeleton and disclination networks}

Frank and Kasper showed that the structure complexity  can be encoded by a network of lines, called by them the ``major skeleton'',  avoiding the $Z_{12}$ sites, and joining the other atomic sites through the Voronoi cells hexagonal faces. The edges of these networks are precisely those edges in the full structure simplicial decomposition sharing six tetrahedral cells. As a whole, the edge set of a F-K structure is therefore a subtle mixture of edges sharing five tetrahedra, and this network of edges sharing six tetrahedra.

F-K phases provide an intermediate solution for the frustrated problem of sphere packing maximal compactness. Best packing are  well known to be given by fcc or hcp packings, with interstices not limited to tetrahedral ones, while a search for a better, but only local, solution would invite to attempt for a maximization of tetrahedral interstices.  The difficulty comes from the fact that a perfect space tiling by regular tetrahedra is impossible, since the dihedral angle of a regular tetrahedron ($cos^{-1}(1/3$) is not a submultiple à $2\pi$, although being close to $2\pi/5$): the problem  is said to be geometrically frustrated.
Icosahedral local order is naturally generated in polytetrahedral packing, but
a perfect packing of regular tetrahedra is only possible in a positively curved space $S^3$ \cite{sadoc1981,sadocmosseribook}. A polytetrahedral packing in $R^3$ will necessarily contain (curvature carrying) topological defects, like disclination lines, corresponding to regions in space where the local order differs from a perfect icosahedral one. Starting with an edge sharing five tetrahedra, the simplest disclination line along this edge will place six tetrahedra around this edge. This corresponds to the F-K major skeleton rules, the latter being therefore identified as such a disclination network\cite{sadocmosseri1982,sadoc1983,nelson1983,sadocmosseri1984}. This network has two types of sites : (i) edge sites, made of $Z_{14}$ polyhedra,  threaded by the edge through opposite points, and (ii) vertex sites, where disclinated edges meet by three at $2\pi/3$ on $Z_{15}$ sites, of by four at  a tetrahedral angle on $Z_{16}$ sites. One can show that these disclination lines cannot be interrupted in the structure: they run throughout the volume, and possibly connect to other lines.

One can distinguish several families of Frank-Kasper phases. A first main difference refers to whether or not atoms can be gathered into simple planes (called, in the F-K terminology, ``primary layers", tiled with triangles, hexagons or pentagons, and ``secondary layers'', with squares and/or triangles), see for instance \cite{sullivan}\cite{sikiric}. We shall describe here the atomic positions, and disclination network, by focusing on the three types of tile interfaces in the secondary layers, either triangle-triangle, square-square and triangle-square (see fig.~\ref{fig:discli_z15_z16}).

 A nice example of a non-layered F-K phase is provided by the Bergman structure \cite{bergman57}, which shows, around symmetrical sites, an interesting nested sequence of clusters with icosahedral symmetry. Among the layered F-K phases, another criterion distinguishes between principal layers containing only hexagons and triangles (like  $A15$, $Z$ and $\sigma$ phases \cite{shoemaker}), and those in which there are also pentagons (leading to the presence of $Z_{16}$ sites, as discussed below, like in the $C15$ and $C14$ Laves phases \cite{friauf,laves}. We shall mainly be interested here to extend these F-K layered phases to quasiperiodic order.

\begin{figure}[htbp]
\includegraphics [width=12cm]{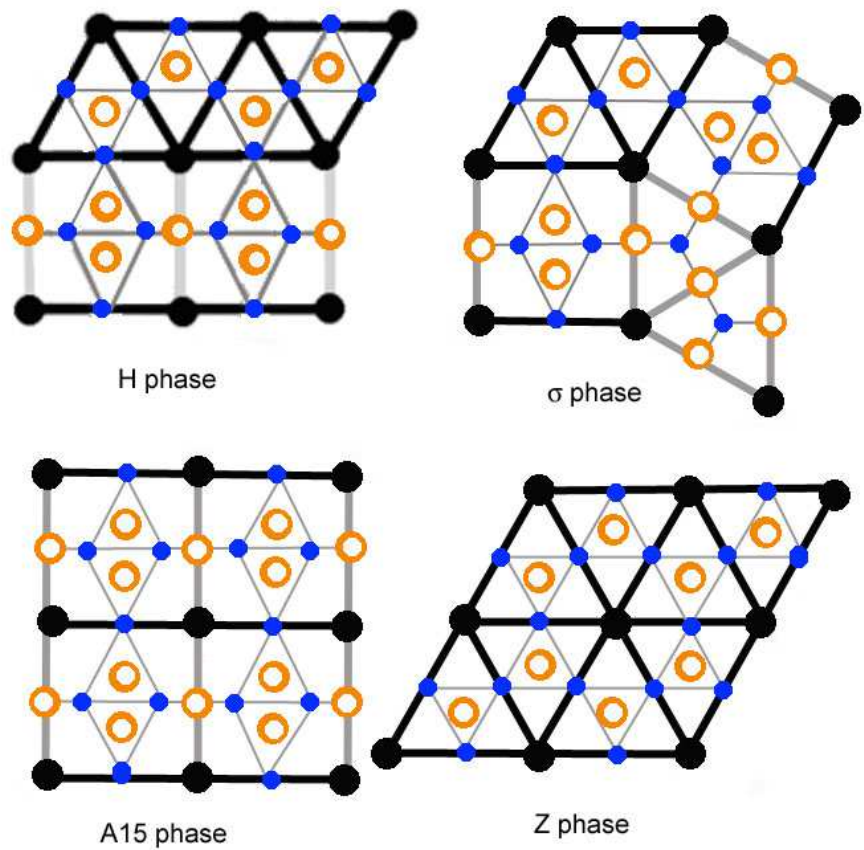}
\caption{\label{fig:decoz15} Atomic decoration of the different layers, illustrated for some classical Frank-Kasper phases, along the two ``primary layers" at heights $0$ and $1/2$ and  the two ``secondary layers'' at heights $1/4$ and $3/4$, the latter two carrying parallel tilings by squares and triangles (or even solely squares or triangles) and being such that atoms are placed at the tile vertices. As explained in the text, there are two sets of edges (marked as dark and light grey edges), with respect to their angular value (modulo $\pi/3$).
Atomic positions on the secondary layers are shown as black dots and atomic positions on the primary layer at height $1/2$ are represented as small blue dots.  As a whole, the atomic positions on this primary layer at height $1/2$ are vertices of a tiling by hexagons and  triangles, with the hexagon centres on top of the atomic positions of the secondary layer at height $1/4$.
The situation is similar on the other primary layer at height $0$, but with decoration rules being reversed. The atomic positions on this layer are marked with red open circles. The full structure is generated as  a periodic repetition of these four layers (the height are given in terms of the square and triangle edge lengths). }
\end{figure}

\subsection{Layered Frank-Kasper phases with only $Z_{12}$, $Z_{14}$ and $Z_{15}$ sites}

Let us now describe more in details the atomic positions in the four equally spaced different layers, which are then periodically repeated. Two "primary" layers are positionned at height $0$ and $1/2$, and two "secondary" layers at height at height $1/4$ and $3/4$. It is convenient, following J. Sullivan\cite{sullivan}, to start  by considering a secondary layer, say at height $1/4$, which is a tiling of squares and/or equilateral triangles (the above heigths are given in terms of the tiles edge length). This tiling plays a central role since it determines the whole atomic structure, up to a decoration choice that will allow or not the presence of $Z_{16}$ sites. In both cases, atoms (or micelles) are located at the vertices of the tiles in this secondary layer, as well as for the parallel other secondary layer located at height $3/4$. The atomic decoration of the two primary layers is case dependant. Finally, the full structure is obtained as a periodic repetition of this four-layer stacking.

The decoration details are given in fig.~\ref{fig:discli_z15_z16}-a,-b,-c, and in the caption, because it is more easily understood with an eye on the drawing. Let us make the following remark on the edges of a square-triangle tiling, which help understanding the decoration procedure : With edges orientation measured with respect to one arbitrary edge, all remaining edges fall into two sets regarding their orientation modulo $\pi/3$. Edges of a given triangle will belong to one of the two sets, while adjacent edges of a square will alternate among the two sets. The two types of edges are marked respectively with dark and light colours in fig.~\ref{fig:decoz15}. Primary layers atomic decoration depends on the underlying edge and triangle colouring on the secondary layer, which we shall therefore call decoration rules $I$ and $II$. In summary,
the atomic decoration add atoms on each layer. Those belonging to the two parallel secondary layers sit on the square-triangle tiling vertices. Those on the primary layers eventually form horizontal tilings of  triangles and hexagons.

With these decorations rules,  many well known F-K phases are recovered, as illustrated in fig.~\ref{fig:decoz15} for the so-called $A15$, $H$, $Z$ and $\sigma$ phases. A site  on a secondary layer has six first-neighbours on each of its two surrounding primary layers , plus two additional neighbors located on its two neighbouring secondary layers at distance $1/2$; so this point has coordination $z=14$. These $Z_{14}$ sites lie on vertical disclination lines orthogonal to the layers. Sites on primary layers falling, by projection, onto (triangle or square) edge centres (smaller blue dots on dark gray edges or open circles on light gray edges) have an icosahedral coordination ($Z_{12}$). Points falling onto triangle centers have $z=15$; these $Z_{15}$ sites belong to three half disclination lines lying on  a primary plane. Finally, the atomic sites falling entirely inside squares are $Z_{14}$ sites located on horizontal disclination lines.

Since this decoration is unique, it is possible to compute the mean coordination number and express the phase composition in terms of the different  types of sites. For that purpose, we express a three-dimensional vector $\mathcal{N}$ with components $(n_{12}, n_{14}, n_{15})$, the number of $Z_{12}$, $Z_{14}$ and $Z_{15}$ sites in terms of a  vector $\nu$ with components the number of squares and triangles. The above described atomic decoration (see fig.~\ref{fig:discli_z15_z16}) can be encoded in the following relation, with a rectangular matrix:

\begin{equation}
\mathcal{N}=\left(
            \begin{array}{cc}
              2 & 3/2 \\
              6 & 1 \\
              0 & 1 \\
            \end{array}
          \right)\nu
\end{equation}

With $y = n_s/n_t$, the average coordination number for the decorated structures reads :$\bar{z}=(108 y+47)/(8 y +7/2)$.
The $A15$ phase has only squares in the secondary layer (therefore with infinite $y$), leading to $\bar{z} = 13.5$. The $Z$ phase has only triangles ($y = 0$), leading to $\bar{z} \simeq 13.429$, and the
 $\sigma$ phase corresponds to $y = 1/2$, leading to $\bar{z} \simeq 13.467$

%
\begin{figure}[tbp]
\includegraphics [width=12cm]{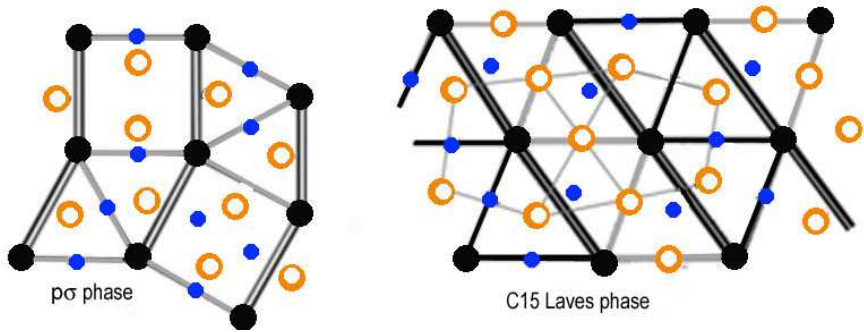}

\caption{\label{fig:decoz16} Decoration of layers in some  Frank-Kasper phases containing $Z16$ sites. As explained in the text, the triangle-square tiling now carry stripes delimited by double edges, and forming closed loops or ending at the boundary. For a given square-triangle tiling, there are many ways to draw these stripes, even if one aims to maximize their numbers (or, said differently, to minimize their individual width). One point should be stressed here: while for the $Z_{16}$-free F.K. layered structures, the secondary layer tiling is done with perfect squares and equilateral triangles, the atomic positions in the present case will relax in such a way that squares will often get distorted into rectangles and triangles turned into being simply isosceles.  Notice finally that, with this decoration,  sites at the triangle-square tiling vertices  on the secondary layers are now $Z_{12}$ sites when belonging to double edges (which therefore concerns all these sites  when a dense set of double edges is used).
}

\end{figure}

\subsection{Layered Frank-Kasper phases containing $Z_{16}$ sites}

We now describe a more complex decoration procedure which allows for the existence of $Z_{16}$ sites. When the former procedure gave a unique atomic decoration on top of a square-triangle tiling, the present one leads to a large number of possible atomic decorations on top of a given triangle-square tiling. While in the $Z_{16}$-free case, the disclination segments are either in or   orthogonal to the layers,  the  main idea here  consists in introducing disclination segments connecting the layers. In order to do that, one defines stripes covering the triangle-square tiling, and delimited by  edges which we shall call (and draw as, in the figures) double edges \cite{shoemaker,sikiric}. The rule to follow is that no triangle should have two double edges and no squares two consecutive double edges. The next step is to decide that the previous decoration rules ($I$ and $II$) associated with coloured triangles, is now flipped while crossing neighbouring stripes. As shown on fig.~\ref{fig:discli_z15_z16}-c,-d,-e, this leads, compared to the $Z_{16}$-free case,   to  $Z_{15}$ (resp.  $Z_{14}$) site changed into  $Z_{16}$ (resp.  $Z_{15}$) sites.

With these decorations rules,  other well-known F-K phases are recovered, as illustrated in fig.~\ref{fig:decoz16} for the $p\sigma$, and $C15$ phases. Note that, although the decoration procedure was done on top of a tiling with equilateral triangles and squares, the actual atomic positions  in these F-K  phases display slightly distorted tiles, but without changing the above described disclination network topology.

When the double edge limited stripes are of minimal width, it is again possible to compute the mean coordination number and express the phase composition in terms of the different  types of sites. We express a four-dimensional vector $\mathcal{N}$ with components $(n_{12}, n_{14}, n_{15}, n_{16})$ in terms of the  vector $\nu$ with components the number of squares and triangles. The above atomic decoration is now encoded in the following relation, with a rectangular matrix:

\begin{equation}
\mathcal{N}=\left(
            \begin{array}{cc}
              3 & 2 \\
              2 & 0 \\
              2 & 0 \\
              0 & 1 \\
            \end{array}
          \right)\nu
\end{equation}

With $y = n_s/n_t$, the average coordination number for the decorated structures reads now :$\bar{z}=(40+94y)/(3+7y)$.
This formula applies to the $Z_{16}$-containing Frank-Kasper phases. As an example, Laves phase have only triangles in the secondary layer (therefore with infinite $y$), leading to $\bar{z} = 40/3$. An interesting case is provided by the  Z phase. We have seen above that it can be obtained by decoration on top of a triangular tiling, leading to a structure with $Z_{12}$ , $Z_{14}$ and $Z_{15}$ sites. But this phase can also be generated using the present rule, applied on an underlying square tiling (in which case only $Z_{15}$ sites are generated accross double edges). With $y$ infinite in the previous formula, the Z phase $\bar{z} = 13.429$ averaged coordination number is recovered.

\section{Dodecagonal square-triangle quasiperiodic tiling}

Dodecagonal square-triangle quasiperiodic tilings, first discussed by Stampfli \cite{stampfli}, can be generated along two standard methods :

(i) Tilings with self-similar symmetry which can be constructed directly in two dimensions, with a step by step  division (called inflation/deflation method) of the basic tiles into smaller (but similar) tiles, followed  by a global rescaling. The quasiperiodic tiling is obtained by repeating the inflation and decomposition process ad infinitum.  We use here this inflation scheme, for which a decoration procedure has initially been given by Schlottmann, and detailed for instance in ref.\cite{hermisson}, in terms of five basic proto-tile decorations, and for which we propose below a simplified algorithm leading to the same tilings. In addition, an even simpler rule in that procedure will allow to generate a six-fold symmetric quasiperiodic tiling.

(ii) The "cut and project" approach. One starts from a periodic structure defined in a higher $4-$dimensional space, with a suitable decomposition into a ``physical" $2-$dimensional space and a $2-$dimensional ``perpendicular" space; a selection step (selection of points projected into an ``acceptance domain" in the ``perpendicular space") is then processed followed by a projection onto the physical space. This method is quite generic, allowing for example a simple computation of the Fourier space structure factor, and a natural definition of the large-cell periodic approximants. However the cut and project method is here operationnally quite complex because, in the perpendicular space, the acceptance domain has a fractal shape \cite{baake,frettloh}. We shall therefore not explicitely use it here, except when showing the acceptance domains in the figures.

%
%
%
\begin{figure}[tbp]
\includegraphics {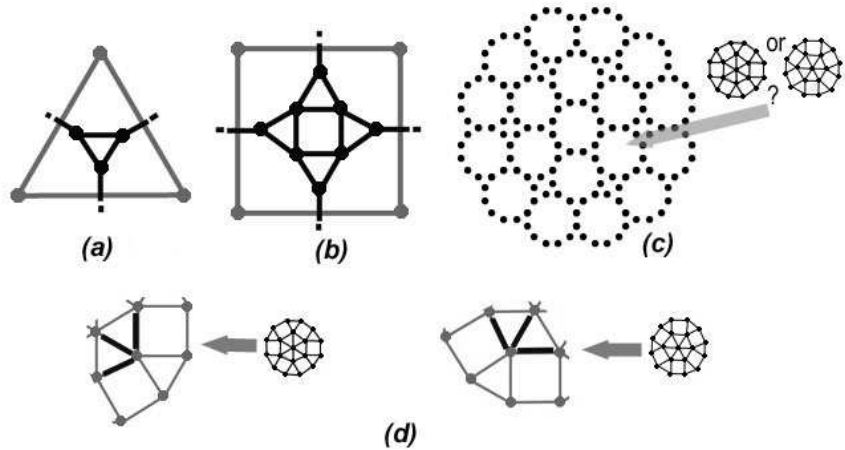}

\caption{\label{fig:dodecarule} Iterative decoration of tiles for a self similar dodecagonal triangle-square tiling: (a) Triangle decoration; (b) square decoration; (c) point set obtained after decoration, forming a tiling of dodecagons (surrounding the vertices of  the previous iteration), squares and triangles (only vertices are represented); There are two possible decorations for the docedagons, as shown in the drawing; (d) the rule for choosing the dodecagon decoration. At a given step in the 12-fold symmetric tiling, there is a small set of site environments.  In order to decorate a given dodecagon : (i)go back to the site located, in the previous iteration, at the dodecagon centre, (ii) determine the edge orientation (modulo $\pi/3$ which is dominant (this is unambiguous), (iii) use this orientation for the internal triangular shell inside the dodecagon. This is done easilly, except for some points on the boundary for which this counting cannot be done due to missing edges. But the dodecagon to be filled on the boundary is similar to an (already decorated) interior dodecagon, to which it is related by the tiling self-similarity.  The rule is then to get the missing information from this interior dodecagon.}
\end{figure}
%

\subsection{Construction by decoration}

We are interested here in generating a 12-fold symmetric triangle-square tiling with a self-similar property around its centre. For that purpose we propose an alternative decoration method, which we believe operationally simple, leading to the same tiling as if obtained with the more standard method proposed by Schlottman. For that purpose we start from a particular seed, with 6 equilateral triangles around the origin.
A simple decoration rule, applied to the triangles (and later to the squares when they will be generated) is shown on fig.~\ref{fig:dodecarule}-a,-b, leading at the next step to a new tiling by dodecagons, squares and triangles with an edge length reduced by a factor $(2-\sqrt{3})$.
The next step, to get a tiling with only triangle and squares, is non trivial. Each dodecagon will be decorated by a first shell made of 6 triangles (an orientation choice is to be made here), and, as a consequence, by a second shell where squares and triangles alternate.

In fact, we can adopt here different strategies, leading to either a disordered random tiling, a 6-fold symmetric tiling (discussed below), or have a precise rule leading to a perfectly self similar 12-fold symmetric tiling. In that case, the orientation degeneracy is lifted by a simple rule, exposed in fig.~\ref{fig:dodecarule}-d, which amounts to a kind of majority rule applied on the edge orientations before decoration.
There are five or six edges meeting at a vertex at iteration $i$. Among these edges, at least three have the same direction modulo $\pi /3$. The rule is then that the decorated dodecagon surrounding this vertex at the next iteration is decorated such as to have its six central edges oriented with this direction modulo $\pi/3$.
 There remains ambiguities for sites on the boundary of the construction, which are solved by taking into account the tiling self-similarity.

Looking at the decoration procedure, and introducing a vector $\nu_i$ whose components are the number of squares and the number of triangles at iteration $i$, the vector $\nu_{i+1}$ is obtained from a transfer matrix as:
\begin{equation}
\nu_{i+1}=\left(
            \begin{array}{cc}
              7 & 3 \\
              16 & 7 \\
            \end{array}
          \right)\nu_{i}
          \label{equa1}
\end{equation}

Some informations about the asymptotic tiling are readily obtained from the Perron root of the matrix, $7+4\sqrt{3}$, and its associated eigenvector $(\sqrt{3}/4,1)$, which give the asymptotic ratio $\sqrt{3}/4$ between the number of squares and triangles (which already rules out the possibility for this tiling to be periodic). Notice however that the area covered by squares equals that covered by triangles.

%
%
\begin{figure}[tbp]
\includegraphics  {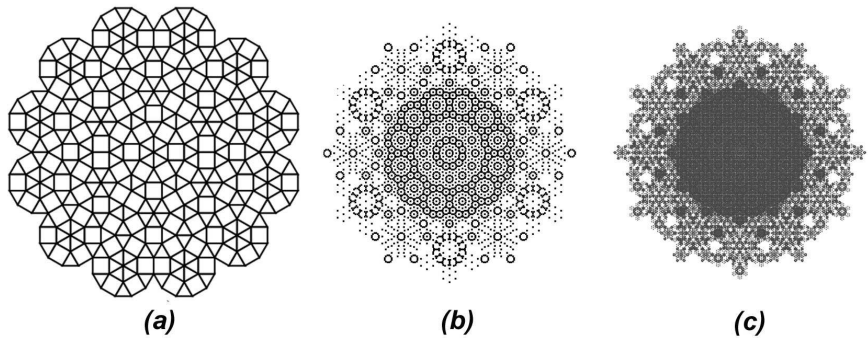}

\caption{\label{fig:dodec_tiling_and_za}a) A piece of a 12-gonal symmetric square-triangle tiling obtained upon two iterations from a dodecagonal seed; b) the associated fractal-like acceptance domain if constructed instead by a cut and project algorithm, showing at this step a 6-fold summetry; c) the asymptotic 12-fold symmetric fractal domain reached after a third iteration }
\end{figure}
%

Figure~\ref{fig:dodec_tiling_and_za} represents the tiling obtained using this rule starting from a decorated dodecagon after two iterations. Also shown is the approximative acceptance domain (in the cut and project approach) that would be reached if this tiling had been constructed along this method. At this step the symmetry is only 6-fold (it is strictly 6-fold in the tiling).  But upon iteration, the acceptance domain turns into a nice fractal shape with 12-fold symmetry.

%
\begin{figure}[htbp]
\includegraphics{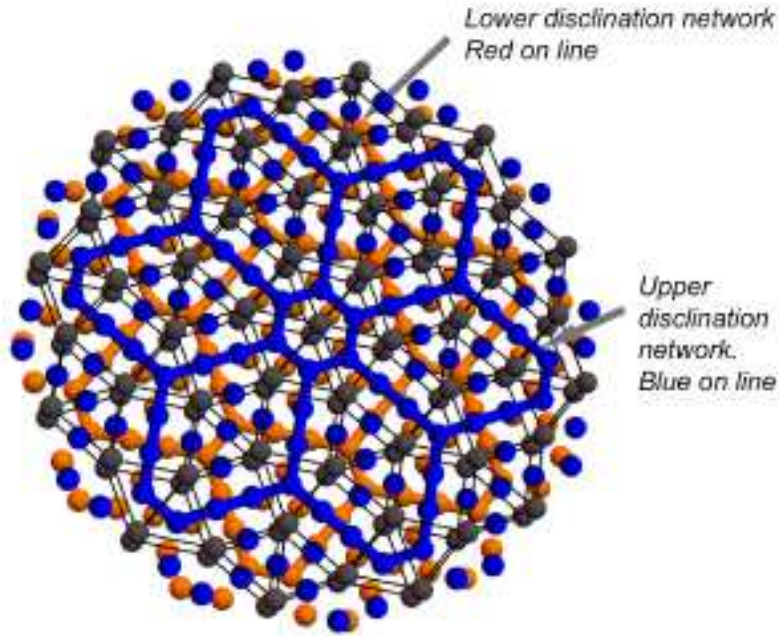}

\caption{\label{fig:dodecFKz15} Frank-Kasper decoration of a dodecagonal quasicrystal. Network of disclinations are represented inside two layers, respectively in red and blue (on-line). There are also disclinations (not drawn here) orthogonal to the layers, threading grey sites located at the triangle-square tiling vertices.}
\end{figure}

%

\section{Quasiperiodic Frank-Kasper phases generated from the square-triangle quasiperiodic tilings}

\subsection{Quasiperiodic Frank-Kasper phases with only $Z_{12}$, $Z_{14}$ and $Z_{15}$ sites}

Once the quasiperiodic 12-fold symmetric triangle-square tiling has been constructed, it becomes a simple exercise to use it as a template and build a generalized quasicrystalline F.K. phase, which is quasiperiodic along the layers, and periodic along the third direction. For the $Z_{16}$-free structure, the obtained structure is unique, and is displayed on fig.\ref{fig:dodecFKz15}, together with a piece of the disclination network (this structure was already briefly discussed in \cite{mosserisadoccras}). The disclination network is made of several disconnected parts. Sites on the secondary layers are $Z_{14}$ sites located at the triangle-square tiling vertex positions, leading therefore to vertical disclinations with this geometry. Each primary layer contains a disclination network having the form  of a particular hexagonal tiling with $Z_{15}$ sites at its nodes and $Z_{14}$ sites, grouped by pairs, along some (long) edges. Sites in the primary layer at height $1/2$ are shown in blue (colour on-line) in fig.\ref{fig:dodecFKz15}, with those (blue) sites not belonging to the disclination network being  $Z_{12}$ sites. The situation is similar on the primary layer at height $0$, with atomic sites shown in red.

From the above computed Perron eigenvector associated with the quasiperiodic triangle-square decoration, one gets the asymptotic square to triangle ratio, $y= \sqrt{3}/4$. The average coordination for the quasicrystal is therefore $\bar{z}=13.464\dots$ and we can extract an (unnormalized) composition, in terms of the coordination numbers~: $Z_{12}^rZ_{14}^sZ_{15}^t$ with $r=3+\sqrt{3}$, $s=2+3\sqrt{3}$ and $t=2$.

%
\begin{figure}[htbp]
\includegraphics{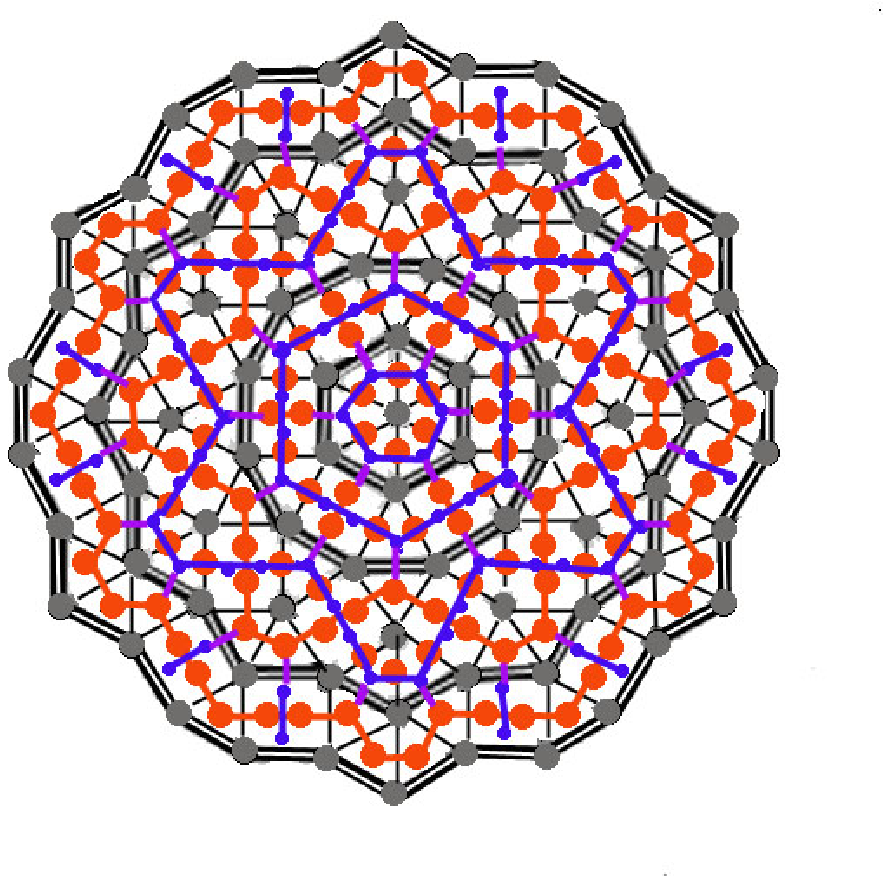}

\caption{\label{fig:dodecFKz16} Frank-Kasper decoration of a dodecagonal quasicrystal with the rule inducing $Z_{16}$ sites. The decoration consists first in drawing stripes delimited by double edges, where  the decoration rules swap and the disclination networks connect from a layer to another. This decoration is not unique; we show here an example with a dense array of concentric stripes. The drawing does not show the $Z_{12}$ sites on the primary layers and therefore only displays sites on the disclination networks on these layers : the $Z_{14}$ sites, grey, in the secondary layers, and $Z_{14}$ and $Z_{15}$ sites in the primary layers. Grey sites sites on the secondary layers are either $Z_{12}$ sites if belonging to double edges, or $Z_{14}$ sites on simple edges. Disclination networks are drawn in red (resp. in blue), colour on-line, for the primary layer at height $0$ (resp. $1/2$). Disclination segments crossing double edges (and connecting the two primary layers) appear in purple (colour on-line).
}
\end{figure}

%

\subsection{Quasiperiodic Frank-Kasper phases containing $Z_{16}$ sites}

Let us now describe quasiperiodic structures containing $Z_{16}$ sites. As discussed above in the crystalline case, the rule consists in defining stripes delimited by double edges, which form the locus where disclination networks connect between different layers. But, while simple dense periodic arrays of stripes were easy to define in the former case, this task is more complex here, and there are a priori multiple solutions. Here, we display one such solution which consists in having a dense array of concentric stripes around the tiling centre (see fig. \ref{fig:dodecFKz16}).

\section{Concluding remarks}

Triangle-square tilings  are known to provide very interesting templates for Frank-Kasper tetrahedrally close packed structures. We have given a detailed description of the decoration procedure which connect them, and focused in addition on the resulting disclination network geometry. Two types of decorations are discussed, leading, or not, to structures with $Z_{16}$ sites.

The Frank-Kasper construction can be generalized to quasicrystalline order, as discussed in the second part of this paper for compact structures based on the 12-fold symmetric triangle-square tiling. This is quite interesting in view of a renewed interest due to the experimental  finding of 12-gonal quasicrystalline soft-matter systems.

 We have described how to generate a self-similar such tiling, as well as (in the appendices) a related 6-fold symmetric tiling, and a square-triangle tiling which is quasiperiodic in one direction and periodic in the other. In all cases, the decorations rules discussed in the periodic case can be used and produce quasicrystalline F-K phases with, as expected, a quasiperiodic disclination network.

 It is well known that topological disorder can be introduced in square-triangle tilings, via so-called ``zipper moves" \cite{oxborrow93}. It would certainly be interesting to study these disordered F-K phases more in details


\begin{acknowledgements}
J-F Sadoc is pleased to thank  M. Imperor, A. Jaganathan and B. Pansu for fruitful discussions

\end{acknowledgements}

%
%
\begin{figure}[tbp]
\includegraphics{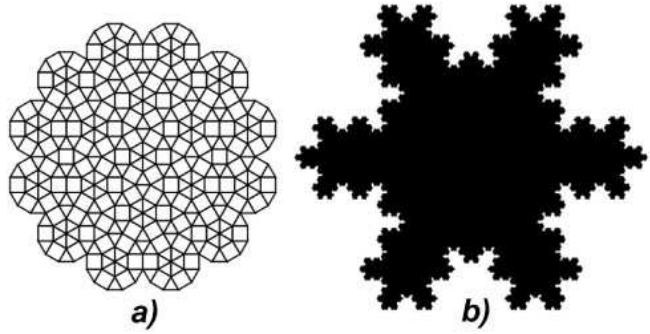}

\caption{\label{fig:quasi6fold}(a) Square-triangle 6-fold symmetric quasiperiodic tiling, with all dodecagon decoration equally oriented (first hierarchical step on top of a decorated dodecagon).  This tiling has an exact 6-fold symmetry axis. (b) Acceptance domain for the same tiling but after a third hierarchical step; the boundary has a fractal shape.}
\end{figure}
%

\section*{Appendix A : A square-triangle quasicrystalline tiling with 6-fold symmetry}
 In order to construct the above 12-fold symmetric square-triangle tiling, we have applied an iterative decoration procedure  inside the dodecagon generated at each step,  with a rule to decide among the two possible orientations modulo $\pi/6$ of the six central triangles. But we can also decide to decorate these dodecagons with a unique orientation. As a consequence, a six-fold symmetric quasiperiodic tiling is generated, as displayed on the figure~\ref{fig:quasi6fold}-a with one direction choice. Interestingly, the acceptance domain, shown  on figure~\ref{fig:quasi6fold}-b, is less complex than that used for the 12-fold case, but still has a fractal boundary (whose dimension can be calculated in that case).

 \begin{figure}[h]
\includegraphics [width=10cm]{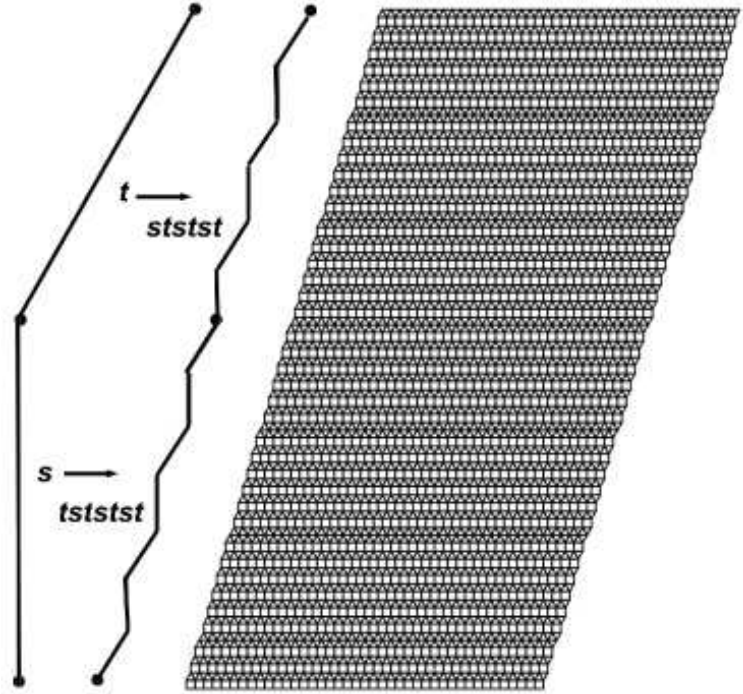}

\caption{\label{fig:quasi1d}
A tiling by rows of squares and triangles, which is quasiperiodic in the direction perpendicular to the rows. It is obtained by a substitution rule on two letters $s$ and $ t$, respectively for rows of squares and triangles, which reads : $t\rightarrow ststst$ and $s\rightarrow tststst$, corresponding to $a=3$ in the matrix given in the text. Recall that this matrix has been chosen in order to have an equal area covered by both squares and triangles, a characteristics which already rules out periodicity. The figure corresponds to three iterations starting from a row of triangles and a row of squares, schematized on the left.}
\end{figure}

 \section*{Appendix B: Square-triangle tiling periodic in one dimension only}

 Interesting new structures can be defined, which interpolate between the F-K $Z$ phase, whose underlying tiling is a triangular one, and the $A15$ phase with a square underlying tiling. It amounts to have linear rows of squares and of triangles, which alternate with any sequence and relative densities. This set is periodic along two directions, with, along the third one, a type of order given by the considered sequence, which can take any form (periodic, quasiperiodic, disordered).
We now describe an explicit such square-triangle tiling, periodic in one direction and quasiperiodic in the orthogonal direction, which also shares with the above 12-fold symmetric tiling the characteristics of equal area covered by the squares and the triangles. We consider a structure made of two types of rows : an infinite row of squares (with unit edge length), labelled $s$, and an infinite row of triangles, of width $\sqrt{3}/2$, labelled $t$.  We define a substitution rule on the two letters $s$ and $ t$, such that a square row is replaced by $a$ square rows and $4$ triangle rows, and a triangle row is replaced by $3$ square rows and $a$ triangle rows. This is encoded in the following transfer matrix form to compute the vector $\mu_{i}=\{n_{s,i},n_{t,i}\}$, with $n_{s,i}$ (resp. $n_{t,i}$) the number of square (resp. triangle) rows at the $i^{th}$ iteration :
\begin{equation}
\mu_{i+1}=\left(
            \begin{array}{cc}
              a & 4 \\
              3 & a \\
            \end{array}
          \right)\mu_{i}
          \label{equa2}
\end{equation}
To the Perron root ($ a+2\sqrt{3}$) of this matrix corresponds an eigenvector ${\sqrt{3}/2,1} $; from this, we get an equal area covered by triangles and squares, for any  value of the integer $a$, and  non-periodicity in the direction orthogonal to rows resulting from the irrationality of $\sqrt{3}$. A simple example is shown figure~\ref{fig:quasi1d} corresponding to $a=3$.
A F-K decoration then leads to a structure which can be viewed as a quasiperiodic mixing of the $A15$ and the $Z$ phase.
Notice that $\mu$ differs from the above $\nu$ in that it counts row frequencies instead of tile frequencies. A triangle row contains alternating triangles of opposite directions, and twice as many triangles as squares, per unit row length. Taking this into account, one recovers the same coordination number $\bar{z}=13.464\dots$ as for the above 12-fold F-K quasicrystal.

Notice that  it is also possible, from the rows, to simply define double edges stripes in such a way as to produce $Z_{16}$ sites. In that case, we obtain, using the expression detailed above, an average coordination $\bar{z} \simeq 13.3812$, quite close to the celebrated value for the Coxeter ``statistical honeycomb" \cite{coxeter58}.


\section*{References}

\begin{thebibliography}{10}

\bibitem{frankkasper}  Frank F.C. \&  Kasper J.S. , Acta Cryst. {\bf 11}, 184 (1958)
and Acta Cryst. {\bf 12}, 483 (1959).

\bibitem{ishimasa}Ishimasa T., Nissen H-U. \& Fukano Y. , \emph{ Phys. Rev. Lett.} \textbf{55}, 511-513 (1985).

\bibitem{chen}Chen H., Li D. X. \& Kuo K. H., \emph{ Phys. Rev. Lett.} \textbf{60}, 1645 (1988).

\bibitem{zeng} Zeng X.B., Ungar G., Liu Y.S., Percec V., Dulcey A. E. \& Hobbs J. K., \emph{ Nature} \textbf{428},  157-160 (2004).

\bibitem{stampfli}  Stampfli P.,\emph{ Helv. Phys. Acta} \textbf{59}, 1260-1263 (1986).

\bibitem{hermisson}  Hermisson J., Richard C. \& Baake M.,\emph{ J. Phys. I France} \textbf{7}, 1003-1018 (1997).

\bibitem{gahler88}  G\"{a}hler F., in  Janot C. \&  Dubois J-M. editors,\emph{ Quasicrystalline materials}, proceedings of ILL/CODEST workshop, World Scientific , 272-284 (1988).

\bibitem{mihalwidom} Mihalkovic M. \& Widom M., Mat. Res. Soc. Symp. Proc., \textbf{805}, LL2.3.1-2.3.6 (2004).

\bibitem{shoemaker} Shoemaker C.B \&  Shoemaker D.P.,  Acta
Cryst., {\bf B 28} 2957-2965 (1972).

\bibitem{bergman57}  Bergman G., Waugh J-L-T \& Pauling L., Acta Cryst. \textbf{10}, 254-259 (1957).

\bibitem{ungar}  Ungar G. \&  Zeng	X., \emph{Soft Matter} \textbf{1},  95106  (2005).

\bibitem{hajiw}  Hajiw S.,  Pansu B. \&  Sadoc J-F, \emph{ACS Nano} \textbf{8},  8116-8121 (2015)

\bibitem{sadoc1981}Sadoc J. F., J. Non-Cryst. Solids \textbf{44}, 1 (1981).

\bibitem{sadocmosseribook} Sadoc J-F. \& Mosseri R., in ``Geometrical
Frustration'', Cambridge Univ. Press (1999).

\bibitem{sadocmosseri1982} Sadoc J-F. \& Mosseri R., J. de Physique Colloque , \textbf{C9-43}, 97 (1982).

\bibitem{sadoc1983} Sadoc J-F., J. de Phys. Lett. \textbf{44},  L707 (1981).

\bibitem{nelson1983}Nelson D.,  Phys. Rev. B \textbf{28}, 5515  (1983).

\bibitem{sadocmosseri1984} Sadoc J-F. \& Mosseri R., J. de Physique, \textbf{45}, 1025-1032 (1984).

\bibitem{sullivan} Sullivan J. M.,  in N. Rivier and J.
F. Sadoc, editors, ``Foams and Emulsions'', \textbf {354} of NATO Advanced Science
Institute Series E: Applied Sciences, Kluwer,(1998)  379-402.
or  Proc. Eurofoam (Delft), 111-119 (2000).

\bibitem{sikiric}  Dutour Sikiric M., O. Delgado-Friedrichs O. \& Deza M., Acta Cryst., A \textbf{66}, 602-615 (2010).

\bibitem{friauf} Friauf J. B, Phys. Rev. {\textbf 29}, 35-40 (1927).

\bibitem{laves} Laves F. \&  Witte H.,
{\it Metallwirtsch Metallwiss Metaltech. }{\bf 14}, 645-649 (1937).

\bibitem{baake}  Baake M., Klitzing R. \&  Schlottman, \emph{ Physica A} \textbf{191}, 554-1809 (1992).

\bibitem{frettloh}  Frettl\"{o}h D.,\emph{ Symmetry: Culture and Science} \textbf{22}, 237-246 (2011).

\bibitem{mosserisadoccras}  Mosseri R. \&  Sadoc	J-F.,	C. R. Physique, \textbf{15},  90-99  (2014)

\bibitem{oxborrow93} Oxborrow M. \& Henley C. L., Phys Rev B \textbf{48}, 6966-6998 (1993)

\bibitem{coxeter58}  Coxeter H. S. M., Close packing and froth, Illinois J. Math. \textbf{2}, 746-758  (1958)
%
%
\end{thebibliography}
\end{document}